\documentclass[a4paper,twocolumn,english,aps,prl,superscriptaddress]{revtex4-1}
\usepackage[T1]{fontenc}
\usepackage[latin9]{inputenc}
\usepackage{color}
\usepackage{babel}
\usepackage{bm}
\usepackage{amsmath}
\usepackage{amssymb}
\usepackage{graphicx}
\usepackage[unicode=true,pdfusetitle,
 bookmarks=true,bookmarksnumbered=false,bookmarksopen=false,
 breaklinks=false,pdfborder={0 0 1},backref=false,colorlinks=false]
 {hyperref}
\usepackage{breakurl}

\makeatletter

\usepackage{babel}
\usepackage{babel}

\makeatother

\begin{document}
\title{Topological Phase Transitions in Disordered Electric Quadrupole Insulators}
\author{Chang-An Li }
\affiliation{School of Science, Westlake University, 18 Shilongshan Road, Hangzhou
310024, Zhejiang Province, China}
\affiliation{Institute of Natural Sciences, Westlake Institute for Advanced Study,
18 Shilongshan Road, Hangzhou 310024, Zhejiang Province, China}
\author{Bo Fu}
\affiliation{Department of Physics, The University of Hong Kong, Pokfulam Road,
Hong Kong, China }
\author{Zi-Ang Hu}
\affiliation{Department of Physics, The University of Hong Kong, Pokfulam Road,
Hong Kong, China }
\author{Jian Li }
\email{lijian@westlake.edu.cn}
\affiliation{School of Science, Westlake University, 18 Shilongshan Road, Hangzhou
310024, Zhejiang Province, China}
\affiliation{Institute of Natural Sciences, Westlake Institute for Advanced Study,
18 Shilongshan Road, Hangzhou 310024, Zhejiang Province, China}
\author{Shun-Qing Shen}
\email{sshen@hku.hk}
\affiliation{Department of Physics, The University of Hong Kong, Pokfulam Road,
Hong Kong, China}
\date{\today }
\begin{abstract}
We investigate disorder-driven topological phase transitions
in quantized electric quadrupole insulators. We show that chiral symmetry
can protect the quantization of the quadrupole moment $q_{xy}$, such that the higher-order topological invariant is well-defined even when disorder has broken all crystalline symmetries.
Moreover, nonvanishing  $q_{xy}$ and consequent corner modes can be induced  from a trivial insulating phase by disorder that preserves chiral symmetry.
The critical points of such topological phase transitions are marked by the occurrence of extended boundary states even in the presence of strong disorder.  We provide a systematic characterization of these disorder-driven topological phase transitions from both  bulk and  boundary descriptions.
\end{abstract}
\maketitle
\textit{\textcolor{blue}{Introduction.-}} Disorder is ubiquitous in
condensed matter systems. A wide range of fundamental phenomena, such as the Anderson
localization and the Kondo effect \cite{PALee85rmp,Anderson58pr,Evers08rmp,Sanchez-Palencia10NP,Kondo64ptp,Hewson97cambr}, are
closely related to disordered systems. When disorder is included in the study of topological phases of matter \cite{Kane10rmp,QiXL11rmp}, the surprising phenomenon of topological Anderson insulators will occur  \cite{LiJian09prl,JiangH09prb,GuoHM10prl,Shem14prl,Meier19Science,Stutzer18nature}, which showcases a nontrivial interplay between disorder and topology.
Recently, the concept of topological invariants in solids
has been generalized to higher orders \cite{Benalcazar17Science,BBH17prb,Schindler18SA,Langbehn17prl,Pengy17prb,SongZD17prl,Khalaf18prb,Youyz18prb,Geier18prb,Ezawa18prl,
Franca18prb,Okugawa18prb,WangZJ19prl,Kudo19prl,Park19prl,Lihq20prl,
Trifunovic19prx,Trifunovic20arxiv, petrides20prr,Schindler18NP,Serra-Garcia18nature,Peterson18nature,Imhof18np,XieBY19prl,ChenXD19prl,Nix19NM,Hassan19NPho,Qiy20prl}.
These higher-order topological insulators, like their conventional cousins,
possess boundary states dictated by bulk topological invariants, but only at even lower dimensions than the latter.
Among the higher-order topological phases, the
quantized electric quadrupole insulator (QEQI) is a prototypical one
that features a quantized electric quadrupole moment in the bulk and zero-energy
modes at the corners \cite{Benalcazar17Science,BBH17prb}. From the outset, a QEQI has been considered as a topological crystalline
insulator \cite{FuL11prl,Neupert18springer}, where the quantization of its electric quadrupole moment is protected
by the underlying crystalline symmetries \cite{BBH17prb}. This apparently poses a no-go condition for the existence of any nontrivial effect induced by disorder in such a system, where all the crystalline symmetries are bound to be broken. As such, a systematic study of the disorder effect in QEQIs, especially its resultant topological phase transitions, remains an open problem despite some related efforts \cite{Araki19prb,SuZ19cpb,Lee18commuphys,LiC20prb,Wang2020arxiv}.

\begin{figure}
\includegraphics[width=1\linewidth]{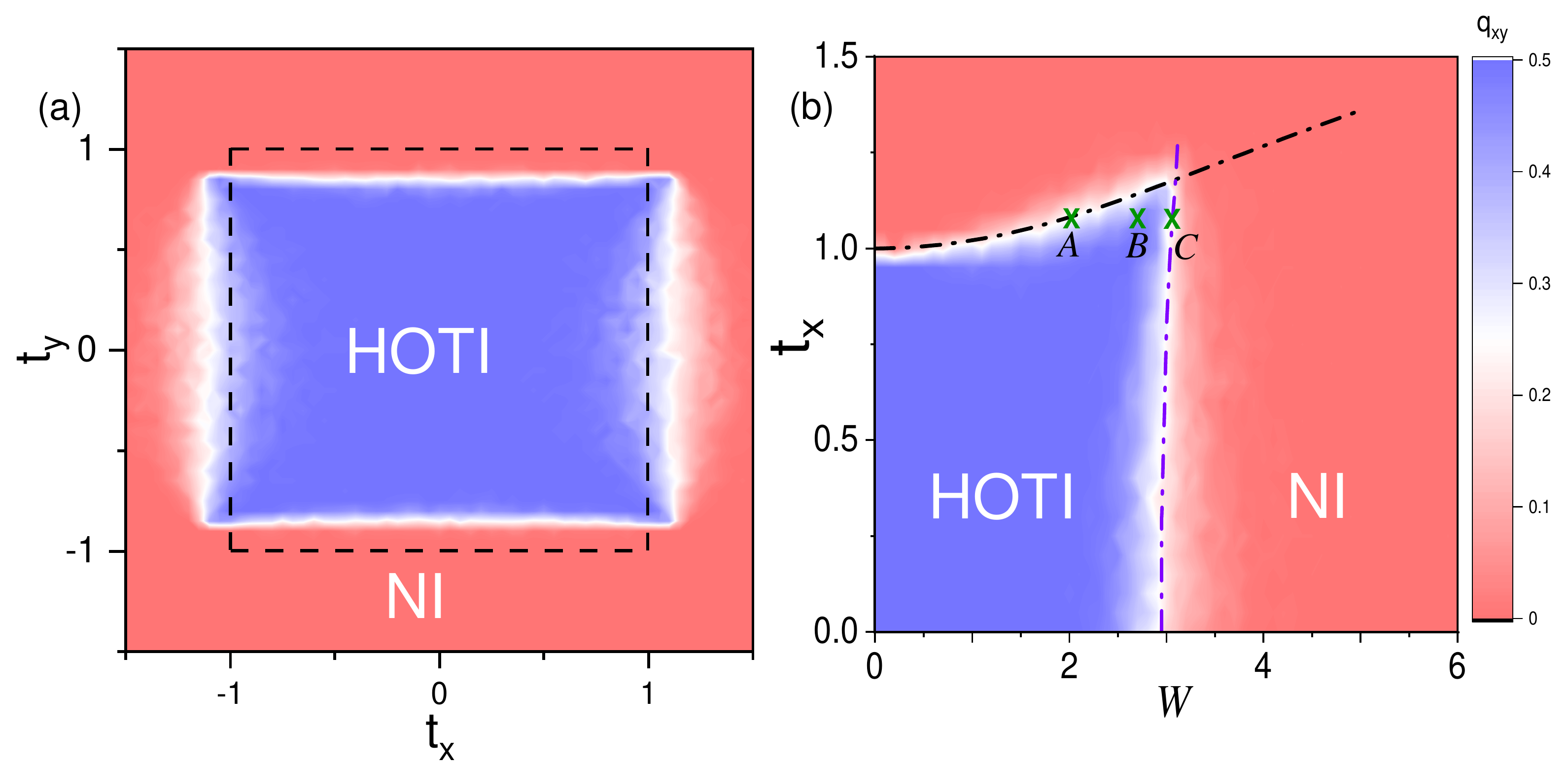}
\caption{Phase diagrams of disordered electric quadrupole insulators. (a) Electric quadrupole moment $q_{xy}$ as a function of mass parameters $t_{x}$ and $t_{y}$ at a fixed disorder strength $W=2.5$. The dashed lines
indicate the phase boundaries in the clean limit. (b) $q_{xy}$ as a function
of $t_{x}$ and $W$ at fixed $t_{y}=0.8$. The dot-dashed lines are the phase boundaries obtained from an effective medium theory. In these phase diagrams, the disorder is of the $V({\bf r})\gamma_{4}$ type, with 120/150 random
configurations averaged in (a)/(b). The system is of size $L_{x}\times L_{y}=30\times30$ with periodical boundary conditions. \label{fig:phase1}}
\end{figure}

In this work, we first prove that the electric quadrupole moment will remain quantized in the presence of disorder, as long as a chiral symmetry is preserved in the system. This allows us to investigate well-defined topological phases in disordered QEQIs. We found that disorder generically introduces a deformation of the phase diagram from the clean limit of a QEQI (see Fig.~\ref{fig:phase1}a). This deformation is nontrivial in the sense that the topological phase regime can expand due to disorder in certain parameter space (see Fig.~\ref{fig:phase1}b). The disordered phase diagrams can be analyzed accurately by using the effective medium theory for the bulk, despite the fact that the topological phase transitions bear no signature in the bulk energy spectrum. Indeed, as an unusual feature of higher-order topological phases, a disorder-induced transition between distinct phases is marked by a localization-delocalization-localization (LDL) transition on specific parts of the system boundary, which leads to a redistribution of fractional charges at the corners of a QEQI. We demonstrate this picture explicitly by combining finite-size scaling analyses with exactly obtained charge densities.

\textit{\textcolor{blue}{Quantized electric quadrupole moments $q_{xy}$ protected
by chiral symmetry.-}}\textcolor{blue}{{} } We consider the following effective Bloch Hamiltonian for a QEQI \cite{Benalcazar17Science,BBH17prb}:
\begin{alignat}{1}
H_{q}({\bf k}) & =t\sin k_{y}\gamma_{1}+[t_{y}+t\cos k_{y}]\gamma_{2}\nonumber \\
 & +t\sin k_{x}\gamma_{3}+[t_{x}+t\cos k_{x}]\gamma_{4},\label{eq:H_q}
\end{alignat}
where the gamma matrices are defined as $\gamma_{j}=-\tau_{2}\sigma_{j}$
{(}$j=1,2,3${)} and $\gamma_{4}=\tau_{1}\sigma_{0}$ with $\tau$
and $\sigma$ both being Pauli matrices but for different
degrees of freedom; $k_{x/y}$ is the wave-vector along $x/y$ (we have set the lattice constant to be unit). The bulk bands of Eq.~\eqref{eq:H_q} are gapped
unless $|t_{x}|=|t_{y}|=|t|$. Without loss of generality, we
will set $t=1$ hereafter. This model respects chiral symmetry $\gamma_{5}^{-1}H_{q}({\bf k})\gamma_{5}=-H_{q}({\bf k})$,
where the chiral symmetry operator $\gamma_{5}\equiv-\gamma_{1}\gamma_{2}\gamma_{3}\gamma_{4}=\tau_{3}\sigma_{0}$.
Since $k_{x}$ and $k_{y}$ are decoupled in Eq.~\eqref{eq:H_q},
the total Hamiltonian can be recast as the sum of two Su-Schrieffer-Heeger
(SSH) models along two directions as $H_{q}(\mathbf{k})=H_{x}(k_{x})+H_{y}(k_{y})$.
In the clean limit, the topologically nontrivial phase is constrained
in the region $|t_{s=x,y}|<1$ where both $H_{x}(k_{x})$ and $H_{y}(k_{y})$
are topologically nontrivial. Under this condition, if an open boundary
with a right-angle corner is considered, we can solve the corner state wavefunction to be of the form $\Psi_{c}(x,y)=\chi_{c}\phi_{x}(x)\phi_{y}(y)$, where $\phi_{x}$ and $\phi_{y}$ are two scalar functions, and $\chi_{c}$ is an eigenstate of the chiral symmetry operator: $\gamma_{5}\chi_{c}=\pm\chi_{c}$
\cite{Li2020}.

When disorder is introduced into the system, the first question we encounter is whether, or when, the electric quadrupole moment will remain quantized, such that a disordered QEQI phase can be well-defined. This question is particularly relevant because QEQIs have been constructed as topological crystalline insulators from the outset, where mirror symmetries are required to ensure the quantization of the electric quadrupole moment. In addition, the nested Wilson loop approach \cite{BBH17prb} originally used to obtain the topological invariant from the momentum space is no longer applicable in the disordered systems. Here, we prove that the quadrupole moment defined in the real space, given by\cite{Kang19prb,Wheeler19prb,Roy19prr}
\begin{equation}
q_{xy}=\frac{1}{2\pi}\mathrm{Im}\,\mathrm{log}\left[\mathrm{det}(U^{\dagger}\hat{Q}U)\sqrt{\mathrm{det}(\hat{Q}^{\dagger})}\right],\label{eq:qxy}
\end{equation}
is indeed quantized even in the presence of disorder as long as the chiral symmetry is preserved. In the above equation:
$\hat{Q}\equiv\exp[i2\pi\hat{q}_{xy}]$ and $\hat{q}_{xy}\equiv\hat{x}\hat{y}/(L_{x}L_{y})$ with $\hat{x}(\hat{y})$ the position operator along the $x(y)$ dimension and $L_{x,y}$ the corresponding size; $\sqrt{\mathrm{det}(\hat{Q}^{\dagger})}=\exp[-i\pi\mathrm{Tr}\hat{q}_{xy}]$; the matrix $U$ is constructed by column-wise packing all the occupied eigenstates, such that $UU^\dag$ is the projector to the occupied subspace.

We sketch our proof as follows and leave the full detail in Supplemental Materials \cite{Li2020}. For $q_{xy}$ to be quantized as an integer multiple of $1/2$, clearly $\mathrm{det}(U^{\dagger}\hat{Q}U)\sqrt{\mathrm{det}\hat{Q}^{\dagger}}$ has to be real. By using Sylvester's determinant identity $\mathrm{det}(\mathbf{1}+AB)=\mathrm{det}(\mathbf{1}+BA)$, and the identity $UU^\dag+VV^{\dagger}=\mathbf{1}$ with $VV^{\dagger}$ the projector to the unoccupied subspace ($V$ is constructed from the unoccupied eigenstates similar to $U$), we obtain $\mathrm{det}(U^{\dagger}\hat{Q}U)=\mathrm{det}(V^{\dagger}\hat{Q}^{\dagger}V)\mathrm{det}\hat{Q}$. Noticing that the chiral symmetry operator relates the occupied states with unoccupied states by $V=\gamma_{5}U$, as well as the fact that $[\gamma_{5}, \hat{Q}]=0$, we have
\begin{equation}
\mathrm{det}(U^{\dagger}\hat{Q}U)=\mathrm{det}(U^{\dagger}\hat{Q}^{\dagger}U)\mathrm{det}\hat{Q}.
\end{equation}
Since $\hat{Q}$ is unitary, it follows immediately that $\mathrm{det}(U^{\dagger}\hat{Q}U)\sqrt{\mathrm{det}Q^{\dagger}}$
is real. In other words, $q_{xy}$ is quantized to be $0$ or $1/2$ as long as the system preserves the chiral symmetry \cite{Li2020}. Note that, in this proof, the explicit form of $\hat{Q}$ is irrelevant except for its commutativity with the chiral symmetry operator, which is generally true because the chiral symmetry is a local symmetry (i.e., diagonal in terms of real-space degrees of freedom) whereas $\hat{Q}$ is constructed from position operators only. Therefore, the conclusion of this proof can be straightforwardly generalized by replacing $\hat{Q}$ with other functions of position operators such as the electric octupole moment operator \cite{Benalcazar17Science,BBH17prb}. In the Supplemental Materials \cite{Li2020}, we further show how to generalize this proof to the case of particle-hole symmetry \cite{Roy19prr}, which is also a local symmetry but does not commute with $\hat{Q}$ because of its anti-unitary nature.

\textit{\textcolor{blue}{Phase diagram of disordered electric quadrupole
insulators}}\textcolor{blue}{.-} With a well-defined topological invariant established for disordered QEQIs, we now present the resulting phase diagrams based on the model in Eq.~\eqref{eq:H_q}. To be specific, we choose one particular type of disorder that preserves the chiral symmetry, represented by $V_{\mathrm{dis}}=V({\bf r})\gamma_{4}$ with the random function $V({\bf r})$
distributed uniformly within the interval $[-W/2,W/2]$ and $W$
being the disorder strength. The averaged quadrupole moment $q_{xy}$ of disordered
QEQIs as a function of $t_{x}$ and $t_{y}$ is shown in Fig. \ref{fig:phase1}(a). Two separate phases can be clearly distinguished: one with $q_{xy}=1/2$ (in blue) signifying a nontrivial higher-order topological insulator (HOTI) phase, and the other with $q_{xy}=0$ (in red) signifying a trivial normal insulator (NI) phase. This phase diagram is more informative when compared with the phase diagram in the clean limit, the phase boundary of which has been marked by dashed lines also in Fig. \ref{fig:phase1}(a). There are obviously contrasting behaviors in terms of the deformation of the HOTI phase regime in the two parameter dimensions ($t_x$ and $t_y$) caused by disorder: the HOTI phase occurs in a shrunk range in $t_y$ but an expanded range in $t_x$ --- the chosen type of disorder is coupled to the same gamma matrix $\gamma_4$ with the latter parameter but not the former one. A more precise analysis of how the deformed phase boundary relies on the disorder type and strength will be given in the next section by employing the effective medium theory and the self-consistent Born approximation (SCBA). Before that, we examine more closely the disorder induced expansion of the nontrivial HOTI phase in the parameter space $t_x$.

In Fig.~\ref{fig:phase1}(b) we show the phase diagram in the $W$-$t_x$ space with fixed $t_y$. We notice again two types of phase boundaries, marked by a black (upper) and a purple (right-side) dot-dashed line, respectively. The upper phase boundary exhibits a clear monotonic increase of the critical $t_x$ with stronger disorder $W$, corresponding to the expanded $t_x$ range by disorder for the HOTI phase in Fig.~\ref{fig:phase1}(a), until it intersects with the right-side phase boundary. As we are set to show in the next section, the right-side phase boundary, which puts an upper bound of the disorder strength for the HOTI phase, originates from the constraint imposed by the disorder-renormalized $t_y$ that has also led to the shrunk range of $t_y$ for the HOTI phase in Fig.~\ref{fig:phase1}(a). These two phase boundaries represent exactly the topological phase transitions that are central to this paper.

\textit{\textcolor{blue}{Effective medium theory of the disorder-induced
topological phase transitions.-}} A better understanding of the disorder-induced topological phase transitions can be achieved with the help of an effective medium theory and the SCBA method \cite{Groth09prl,Park17prb,ChenCZ15prl,LiuS16prl}.
In the SCBA method, the key is to obtain the self-energy introduced by the disorder self-consistently, and then to include the self-energy as renormalization to the original Hamiltonian. In our case, by symmetry arguments the self-energy can be simplified to be
\begin{equation}
\Sigma(E_{F})=\Sigma_{4}\gamma_{4}+\Sigma_{2}\gamma_{2}+\Sigma_{0}I_{4\times4}.
\end{equation}
Specifically, the self-energy $\Sigma$ satisfies the following self-consistent integral
equation
\begin{equation}
\Sigma(E_{F})=\frac{W^{2}}{48\pi^{2}}\int_{BZ}d^{2}{\bf k}\gamma_{4}\frac{1}{E_{F}+i\eta-H_{q}({\bf k})-\Sigma(E_{F})}\gamma_{4},\label{eq:SCBA}
\end{equation}
where the integral runs over the first Brillouin zone, and $\eta$
is an infinitesimal positive number. $E_{F}$ is Fermi energy which is set at zero here, i.e., the system is half filled. From Eq.~\eqref{eq:SCBA}, there are explicitly three coupled self-consistent integral equations that will fully determine $\Sigma$ \cite{Li2020}. After obtaining the self-energy $\Sigma$ (where $\Sigma_{0}$ turns out to be zero at zero energy because of the chiral symmetry), the topological mass terms $t_x$ and $t_y$ are renormalized according to
\begin{subequations}
\begin{align}
&\bar{t}_{x}=t_{x}+\mathrm{Re}\Sigma_{4}, \\
&\bar{t}_{y}=t_{y}+\mathrm{Re}\Sigma_{2}.
\end{align}\label{eq:massren}
\end{subequations}
This produces the new phase boundaries at $|\bar{t}_{x}|=|\bar{t}_{y}|=1$, which formally resemble the conditions in the clean limit but with a key difference in the implicit dependence on $W$.

The preceding approach can quantitatively account for the phase boundaries in the
presence of disorder. The expanded range of $t_x$ and the shrunk range of $t_y$ for the HOTI phase in the disordered phase diagram, as shown in Fig. \ref{fig:phase1}(a), can be understood from the opposite signs of the self-energy contributions $\Sigma_{4}$ and $\Sigma_{2}$, which in turn comes from the anti-commutation relation between $\gamma_{4}$
and $\gamma_{2}$ \cite{Li2020}. In previous discussions, we have seen that only if
the two individual SSH models consisting the full model in Eq.~\eqref{eq:H_q} are topologically nontrivial
simultaneously, the system can possess nontrivial bulk topological
invariant and host zero-energy modes at its corners. Therefore, if
disorder drives one of the two SSH components from
topologically nontrivial to trivial, a phase transition of the higher-dimensional
system will occur. Indeed, the topological phase transitions of the two SSH components are each captured by one of the conditions $|\bar{t}_{x,y}|=1$ with the renormalized mass $\bar{t}_{x,y}({t}_{x,y}, W)$ given by Eq.~\eqref{eq:massren}. For a fixed disorder strength $W$, such as in the case shown in Fig. \ref{fig:phase1}(a), these conditions lead to four critical values of $t_{x}$ and $t_y$, resulting in a rectangular shaped phase boundary in the phase diagram. With varying disorder strength, on the other hand, the two conditions lead directly to the two phase boundary lines demonstrated in Fig. \ref{fig:phase1}(b). Specifically, we plot the solutions to the equations $\bar{t}_x=1$ and $\bar{t}_y=1$ as the dot-dashed lines in black and in purple, respectively. Both lines coincide very well with the phase boundaries obtained from numerically calculating $q_{xy}$ as discussed in previous sections, until the two lines intersect.

\begin{figure}
\includegraphics[width=1\linewidth]{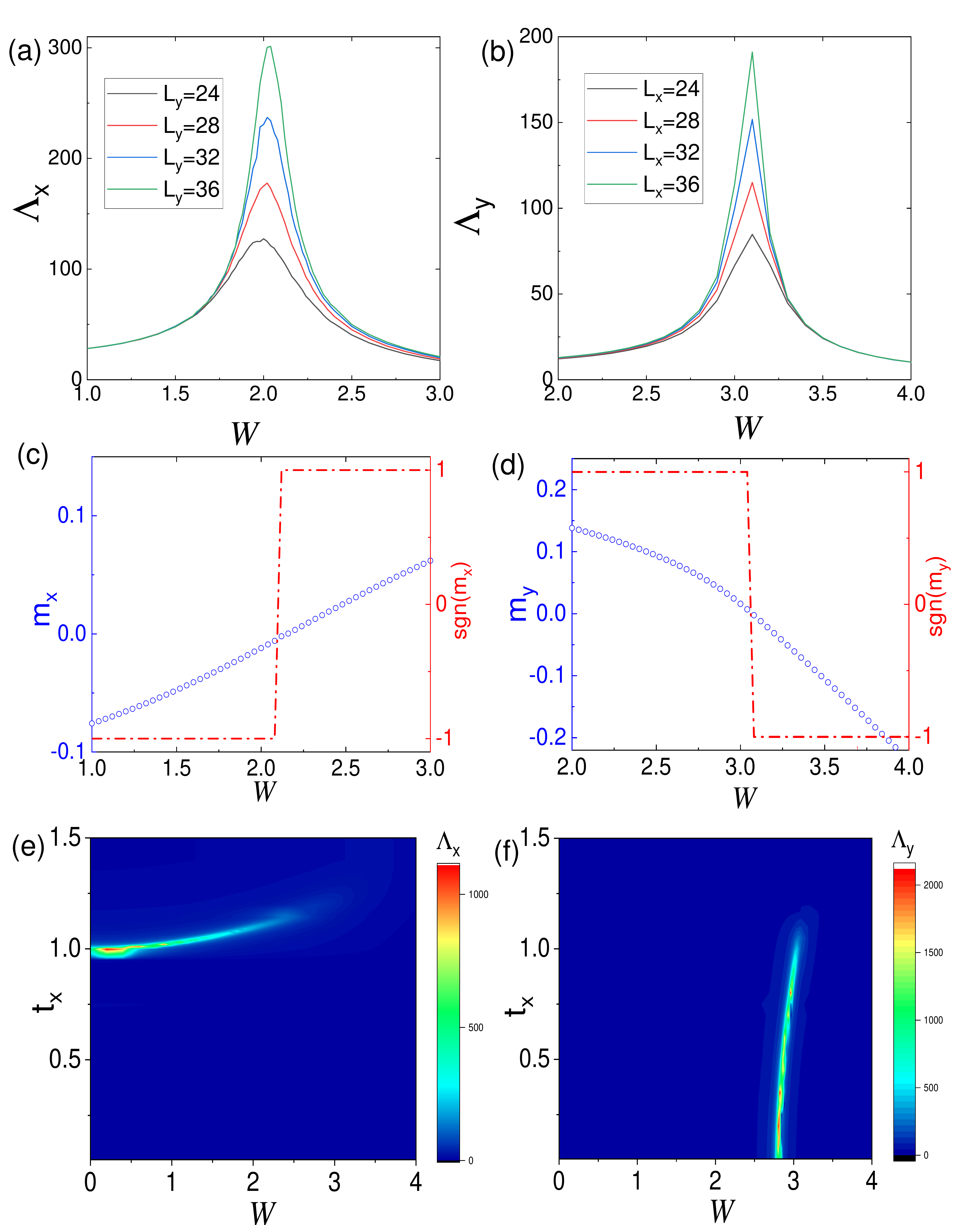}
\caption{Signatures of delocalized states at the open boundaries: a finite-size scaling analysis of the localization
length (a) $\Lambda_{x}$ and (b) $\Lambda_{y}$ as functions of $W$ with an open boundary condition in the transverse dimension; effective boundary mass terms (c) $m_{x}$ and (d) $m_{y}$ as functions of $W$; phase boundaries indicated by the divergence of (e) $\Lambda_{x}$ and (f) $\Lambda_{y}$ as analyzed in the (a) and (b) panels. In panels (a-d), we have set $t_{x}=1.1$ and $t_{y}=0.8$. \label{fig:Criticalpoint}}
\end{figure}

\textit{\textcolor{blue}{Localization-delocalization-localization transitions
at open boundaries.-}} The higher-order topological phase transitions generically show no signatures in terms of the bulk energy spectrum but instead are accompanied by LDL transitions at the open boundaries of the system. To demonstrate this in disordered systems, we perform a finite size scaling analysis based on localization length calculated from the numerical transfer matrix method \cite{MacKinnon83ZP,Kramer93RPP,Yamakage13prb,Li2020}. Specifically, we compare the localization length (at zero energy) along quasi-one-dimensional ribbons of our model system with different width, longitudinal orientations and transverse boundary conditions. The dependence of the localization length on the ribbon width, in a specific orientation and boundary condition setting, indicates the presence or absence of delocalized bulk or boundary states in the thermodynamic limit. We focus on the parameter space close to the phase boundaries identified in the previous sections. With a periodic boundary condition in the transverse dimension, we found that the localization length (upon normalization by the width) decrease monotonically with increasing width in the entire parameter ranges of our interest, regardless of the orientation along the ribbon \cite{Li2020}. This indicates that there is no occurrence of delocalized bulk states during the phase transitions --- the bulk of the system remains insulating. In contrast, when an open boundary condition is considered, the localization length along certain longitudinal orientation can exhibit monotonic increase with increasing width, signifying a divergence in the thermodynamic limit, around a topological phase transition point, as exemplified in Fig.~\ref{fig:Criticalpoint}(a, b). The divergence of the localization length at a critical value of the disorder strength, which occurs only with an open boundary condition and along certain orientation, indicates the existence of delocalized states at the corresponding boundaries, despite strong disorder, at the critical point. We note that the LDL transitions discussed here are similar to the topological phase transitions across Landau levels in the quantum Hall effect \cite{Ando83jpsj}, in the fact that the delocalized states occur only at the exact critical points.

The boundary LDL transitions established above by a finite-size scaling analysis can be further understood with an effective boundary theory \cite{Li2020}, where the (boundary) spectrum around a critical point is controlled by an effective mass, given by $m_{x}=1-t_{x}-\mathrm{\mathrm{Re}[\Sigma_{4}}(W)]$ for the boundaries along $x$, or $m_{y}=1-t_{y}-\mathrm{Re}[\Sigma_{2}(W)]$ for the boundaries along $y$. The critical points are associated with the conditions $m_{x,y}=0$, which coincide with the conditions that we have derived earlier from the bulk phase transitions. By using the SCBA method, the effective mass values and corresponding signs are obtained and shown
in Figs. \ref{fig:Criticalpoint}(c, d), which agree with the finite-size scaling results.

The LDL transitions along each open boundary orientation also enable us to establish the two types of phase boundaries discussed previously in the context of bulk topology. This is shown in Figs.~\ref{fig:Criticalpoint}(e, f) with calculated localization length corresponding to Figs.~\ref{fig:Criticalpoint}(a, b). The full agreement between this approach and the bulk invariant approach manifests the close interconnection between the boundary and the bulk descriptions of the higher-order topological phase transitions.

\begin{figure}
\includegraphics[width=1\linewidth]{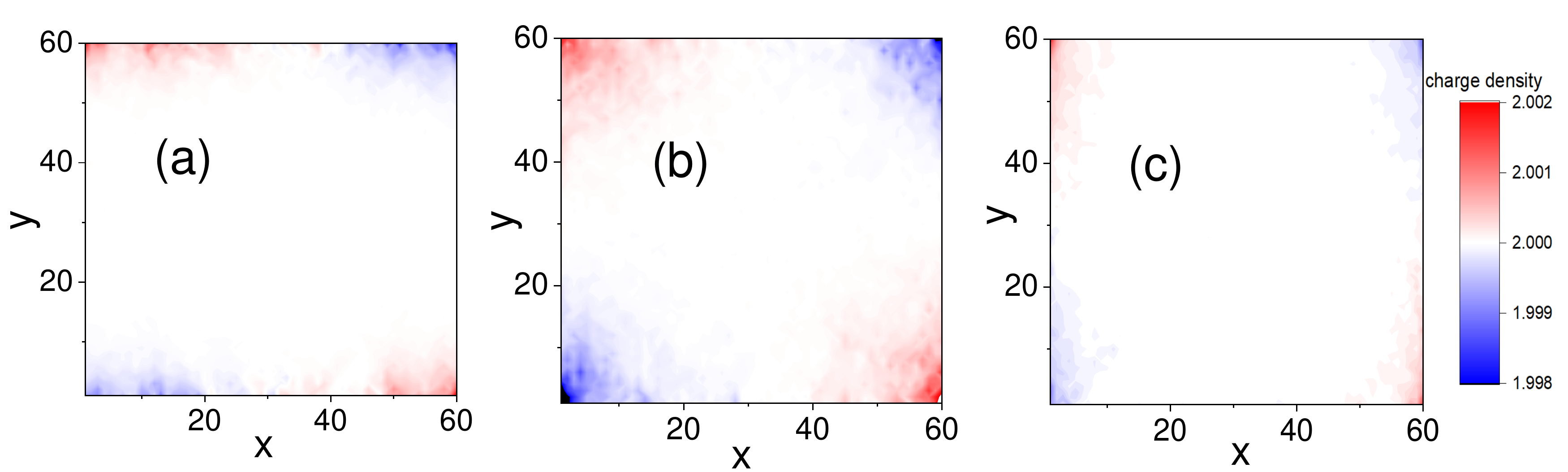}
\caption{Charge density distributions (a, b, c) that correspond to the three points, A, B, and C, (marked by crosses) in the phase diagram in Fig. \ref{fig:phase1}(b). The points $A$ and $C$ sit in different phase boundaries; the point $B$ sits inside the nontrivial QEQI phase. The values of the disorder strength in (a), (b), and (c) are $W=2.02$, $W=2.70$ and $W=3.10$, respectively. In all these calculations we have set $t_{x}=1.1$ and $t_{y}=0.8$, and taken an average of 10 disorder configurations. \label{fig:Cornercharges}}
\end{figure}

\textit{\textcolor{blue}{Charge density redistribution at boundaries and corners.-}}
A hallmark of QEQIs is the presence of fractional charges at the corners which consist in the quantized electric quadrupole. In this section we demonstrate how the disorder-driven topological phase transitions lead to the redistribution of the charge density towards (or away from) the fractional corner charges. For clarity and simplicity, let us focus on three representative points in the phase diagram, marked by $A$, $B$, and $C$ in Fig. \ref{fig:phase1}(b). These three phase points correspond to a fixed $t_{x}$ (we choose $t_{x}=1.1$) but varying disorder strength $W$, such that $A$ and $C$ sit on the two types of phase boundaries respectively, whereas $B$ sits in the nontrivial QEQI phase. The calculated charge densities  for these points are shown in Fig.~\ref{fig:Cornercharges}. At the critical point $A$($C$), the charge density extends only along the $x$($y$) boundaries, as enabled by the occurrence of delocalized states thereat, and exhibits a continuous bipolar form  with opposite polarities (offset by the mean values) on opposite boundaries; at the $B$ point, the charge density displays a more symmetric quadrupolar form that is deformed from the dipoles in $A$or $C$. The charge density in the $B$ point shows clear localization at the corners owing to the topological bulk-corner correspondence, and upon integration over each quadrant sums to the fractional value $\pm 1/2$ with high accuracy when the system size is sufficiently large.

\textit{\textcolor{blue}{Conclusion.-}}
In short, we have presented a comprehensive description of the disorder-induced topological phase transitions in quantized electric quadrupole insulators. It is proved rigorously that the quantization of the electric quadrupole moment $q_{xy}$ can be protected by the chiral symmetry even in the presence of strong disorder. We have also uncovered disorder-driven phase transitions from trivial to higher-order topological phases, which are signified by localization-delocalization-localization transitions at certain open boundaries with the system bulk remaining insulating. We expect this exotic disorder effect can be experimentally demonstrated in, e.g., photonic crystals \cite{XieBY19prl,ChenXD19prl,Hassan19NPho} or electric
circuits \cite{Imhof18np,Peterson18nature} by taking advantage of their high controllability.

C. A. Li thanks Liyuan Chen and A. Weststrom for helpful discussions,
and acknowledges B. Kang and G. Y. Cho for communication on the real
part of quadrupole formula. This work was supported by NSFC under
Grants No. 11774317, NSF of Zhejiang under Grant No. Q20A04005, and the Research Grants Council, University Grants
Committee, Hong Kong under Grants Nos. 17301717 and 17301220. The
numerical calculations were performed on Supercomputer cluster of
Westlake University.

\bibliographystyle{apsrev4-1}

\appendix
\numberwithin{equation}{section}\setcounter{figure}{0}\global\long\def\thefigure{S\arabic{figure}}
\global\long\def\thesection{S\arabic{section}}
\global\long\def\thesubsection{\Alph{subsection}}

\begin{widetext}
\begin{center}
\textbf{\large{}Supplemental materials of ``Topological Phase Transitions in Disordered Electric Quadrupole Insulators''}{\large{} }
\par\end{center}{\large \par}

\section{Proof of the quantization of quadrupole moments protected by chiral symmetry}

In this section, we present the proof that the quantization of quadrupole
moments is protected by chiral symmetry
of the system. The quadrupole moments formula is
\begin{equation}
q_{xy}=\frac{1}{2\pi}\mathrm{Imlog}[\mathrm{det}(U^{\dagger}\hat{Q}U)\sqrt{\mathrm{det}(\hat{Q}^{\dagger})}],
\end{equation}
where the matrix $U$ is constructed by column-wise packing of the
occupied eigenstates, $\hat{Q}=e^{2\pi i\hat{x}\hat{y}/L_{x}L_{y}}$,
and $\hat{x},\hat{y}$ are the position operators. To have quantized
$q_{xy},$ the part in the logarithm function has to be real. Thus
our target is transformed to prove $\mathrm{det}(U^{\dagger}\hat{Q}U)\sqrt{\mathrm{det}(\hat{Q}^{\dagger})}$
is real.

To this end, we perform a deformation of the determinant
as
\begin{alignat}{1}
\mathrm{det}(U^{\dagger}\hat{Q}U) & =\mathrm{det}[U^{\dagger}(\hat{Q}-\mathbf{1}+\mathbf{1})U]\nonumber \\
 & =\mathrm{det}[\mathbf{1}+U^{\dagger}(\hat{Q}-\mathbf{1})U].
\end{alignat}
Using the Sylvester's determinant identity $\mathrm{det}(\mathbf{1}+AB)=\mathrm{det}(\mathbf{1}+BA)$,
the above equation is simplified to be
\begin{alignat}{1}
\mathrm{det}(U^{\dagger}\hat{Q}U) & =\mathrm{det}(\mathbf{1}+(\hat{Q}-\mathbf{1})UU^{\dagger}).
\end{alignat}
Note that $UU^{\dagger}=P_{\mathrm{occ}}$ is the projection operator
projected to occupied states, and $P_{\mathrm{occ}}=\mathbf{1}-VV^{\dagger}$
where $V$ is constructed by the unoccupied states. Thus
\begin{alignat}{1}
\mathrm{det}(U^{\dagger}\hat{Q}U) & =\mathrm{det}[\mathbf{1}+(\hat{Q}-\mathbf{1})(\mathbf{1}-VV^{\dagger})]\nonumber \\
 & =\mathrm{det}[\hat{Q}-(\hat{Q}-\mathbf{1})VV^{\dagger}]\nonumber \\
 & =\mathrm{det}[\mathbf{1}+(\hat{Q}^{\dagger}-\mathbf{1})VV^{\dagger}]\mathrm{det}\hat{Q}\nonumber \\
 & =\mathrm{det}(V^{\dagger}\hat{Q}^{\dagger}V)\mathrm{det}\hat{Q}.
\end{alignat}

Let us focus on the system with chiral symmetry. If the system
respects chiral symmetry, i.e., $\gamma_5^{-1}H\gamma_5=-H$, the occupied states
and unoccupied states are related by chiral symmetry operator as
\begin{equation}
V=\gamma_5U.
\end{equation}
Utilize this relation and note that $[\gamma_5,\hat{Q}]=0$, we have
\begin{alignat}{1}
\mathrm{det}(U^{\dagger}\hat{Q}U) & =\mathrm{det}(V^{\dagger}\hat{Q}^{\dagger}V)\mathrm{det}\hat{Q}\nonumber \\
 & =\mathrm{det}(U^{\dagger}\gamma_5^{\dagger}\hat{Q}^{\dagger}\gamma_5U)\mathrm{det}\hat{Q}\nonumber \\
 & =\mathrm{det}(U^{\dagger}\hat{Q}^{\dagger}U)\mathrm{det}\hat{Q}.
\end{alignat}
As the matrix $\hat{Q}$ is unitary, we have
\begin{alignat}{1}
\mathrm{det}(U^{\dagger}\hat{Q}U) & \sqrt{\mathrm{det}\hat{Q}^{\dagger}}=\mathrm{det}(U^{\dagger}\hat{Q}^{\dagger}U)\sqrt{\mathrm{det}\hat{Q}},
\end{alignat}
 thus
\begin{alignat}{1}
\mathrm{det}(U^{\dagger}\hat{Q}U) & \sqrt{\mathrm{det}\hat{Q}^{\dagger}}=\left(\mathrm{det}(U^{\dagger}\hat{Q}U)\sqrt{\mathrm{det}\hat{Q}^{\dagger}}\right)^{*},
\end{alignat}
Finally, the determinant $\mathrm{det}(U^{\dagger}\hat{Q}U)\sqrt{\mathrm{det}\hat{Q}^{\dagger}}$
is real. It is proved that $q_{xy}$ is quantized to $0$ or $\frac{1}{2}$.

Actually, the above conclusion can be generalized to system with particle-hole symmetry.  The particle-hole symmetry operator is represented
by an anti-unitary operator $P=U_{p}K$ where $U_{p}$ is a unitary matrix and $K$ represents complex conjugation. Thus the occupied states and unoccupied states are related by the particle-hole symmetry
operator as
\begin{equation}
V=PU=U_{p}KU=U_{p}U^{*}.
\end{equation}
Then following the above formula, we have
\begin{alignat}{1}
\mathrm{det}(U^{\dagger}\hat{Q}U) & =\mathrm{det}(U^{T}U_{p}^{\dagger}\hat{Q}^{\dagger}U_{p}U^{*})\mathrm{det}\hat{Q}\nonumber \\
 & =\mathrm{det}(U^{T}\hat{Q}^{\dagger}U^{*})\mathrm{det}\hat{Q}\nonumber \\
 & =\mathrm{det}(U^{\dagger}\hat{Q}^{T}U)^{*}\mathrm{det}\hat{Q}\nonumber \\
 & =\mathrm{det}(U^{\dagger}\hat{Q}U)^{*}\mathrm{det}\hat{Q}.
\end{alignat}
Finally, we still have
\begin{alignat}{1}
\mathrm{det}(U^{\dagger}\hat{Q}U) & \sqrt{\mathrm{det}\hat{Q}^{\dagger}}=\left(\mathrm{det}(U^{\dagger}\hat{Q}U)\sqrt{\mathrm{det}\hat{Q}^{\dagger}}\right)^{*}.
\end{alignat}
It is proved that $q_{xy}$ is also quantized to $0$ or $\frac{1}{2}$
by particle-hole symmetry.

\section{Effective low-energy edge Hamiltonian and corner modes solution}

In this section, we derive the low-energy edge Hamiltonian for the
quantized electric quadrupole insulators. For the chosen parameters, the bulk gap closes at the $(k_{x},k_{y})=(\pi,\pi)$
point when $t_{x}=t_{y}=1$. Expanding Eq.~(1) in the main text around
this gap closing point to the second order, we have an effective model
\begin{equation}
H({\bf k})=k_{y}\gamma_{1}+M_{y}\gamma_{2}+k_{x}\gamma_{3}+M_{x}\gamma_{4},
\end{equation}
where $M_{x,y}\equiv m_{x,y}-bk_{x,y}^{2}$ and $m_{x,y}\equiv1-t_{x,y},$ and
$b=\frac{1}{2}$. Note that an overall minus sign is neglected. This
effective bulk Hamiltonian inherits all symmetries of the original
model.

The target is to get an effective edge Hamiltonian starting from this effective bulk Hamiltonian. To this end, let
us consider a semi-infinite plain $x\geq0$ and keep $k_{y}$ a good
quantum number. The wave function thus has a form $\Psi(x,k_{y})=\psi(x)e^{ik_{y}y}$.
Then the Schrödinger equation $H({\bf k})\Psi(x,k_{y})=E\Psi(x,k_{y})$
leads to
\begin{equation}
\left[-i\partial_{x}\gamma_{3}+M_{x}\gamma_{4}\right]\psi(x)+\left(k_{y}\gamma_{1}+M_{y}\gamma_{2}\right)\psi(x)=E\psi(x),
\end{equation}
where we have replaced $k_{x}\rightarrow-i\partial_{x}$, and $E$
is the energy. The first parentheses has grouped all the dependence
on the $x$ coordinate, and this equation can have solution only if the
first parentheses gives constant. For simplicity we set this constant
to be zero and obtain

\begin{equation}
(-i\partial_{x}+M_{x}\gamma_{3}\gamma_{4})\psi(x)=0.
\end{equation}
Simplifying it further,
\begin{equation}
\partial_{x}\psi(x)=M_{x}\tau_{3}\sigma_{3}\psi(x).
\end{equation}
Thus $\psi(x)$ should be an eigen function of $\tau_{3}\sigma_{3}$
as $\psi(x)=\phi(x)\chi_{\tau,\sigma}$ where $\tau_{3}\sigma_{3}\chi_{\tau,\sigma}=\tau\sigma\chi_{\tau,\sigma}$
with $\tau,\sigma=\pm1$. Taking the trial wave function $\phi(x)=e^{-\zeta x}$,
we have the secular equation
\begin{equation}
b\zeta^{2}-\tau\sigma\zeta+m_{x}=0.
\end{equation}
The two roots are $\zeta_{1,2}=\frac{\tau\sigma\pm\sqrt{1-4m_{x}b}}{2b}$,
and they satisfy the relation

\begin{equation}
\zeta_{1}+\zeta_{2}=\frac{\tau\sigma}{b},\text{\ensuremath{\zeta_{1}\zeta_{2}=\frac{m_{x}}{b}.}}
\end{equation}
To have edge states, the two roots should be positive, which constrains
the condition

\begin{equation}
m_{x}>0,\sigma=\mathrm{sgn}(\tau).
\end{equation}
Under this constraint, then $\chi_{\tau,\sigma}$ is reduced to positive
eigenvalues as $\tau_{3}\sigma_{3}\chi_{\tau,\sigma}=+1\chi_{\tau,\sigma}$.
The spatial part of wave functions are now $\phi(x)=C(e^{-\zeta_{1}x}-e^{-\zeta_{2}x})$
where $C$ is a normalization factor.

Now we have two states $\psi_{+}=\chi_{\tau=1,\sigma=1}\phi(x)$ and
$\psi_{-}=\chi_{\tau=-1,\sigma=-1}\phi(x)$. Then projecting the remaining
part of the Hamiltonian into the subspace spanned by these two states,
we find the effective edge Hamiltonian along the $y$ direction as

\begin{equation}
H_{\mathrm{edge}}(k_{y})=(m_{y}-bk_{y}^{2})\rho_{x}-k_{y}\rho_{y},
\end{equation}
where $\rho_{x,y}$ are the Pauli matrices in the basis $\{\psi_{+},\psi_{-}\}$.
This Hamiltonian describes exactly the one-dimensional Dirac Hamiltonian \cite{SQS}.
It is topologically nontrivial if $\mathrm{sgn}(m_{y}b)>0$ and topologically
trivial if $\mathrm{sgn}(m_{y}b)<0$. This result is consistent with
the phase diagram of quantized electric quadrupole insulators. Take a similar
approach, the effective edge Hamiltonian along the $x$ direction
is

\begin{equation}
H_{\mathrm{edge}}(k_{x})=(m_{x}-bk_{x}^{2})\rho_{x}-k_{x}\rho_{y}.
\end{equation}

Next we present the solution of corner modes staring again from the bulk effective mode. Consider
the zero-energy states at the corner $x,y\geq0$. The corner modes
are assumed to be

\begin{equation}
\Psi_{c}(x,y)=\chi_{c}\phi_{x}(x)\phi_{y}(y),
\end{equation}
which should satisfy the boundary condition
\begin{equation}
\Psi_{c}(x=0,y)=\Psi_{c}(x,y=0)=\Psi_{c}(x=\infty,y=\infty)=0.
\end{equation}
 Then the Schrödinger equation $H({\bf k})\Psi_{c}(x,y)=0$ leads
to
\begin{alignat*}{1}
(-i\partial_{x}+M_{x}\gamma_{3}\gamma_{4})\chi_{c}\phi_{x}(x)=0,\\
(-i\partial_{y}+M_{y}\gamma_{1}\gamma_{2})\chi_{c}\phi_{y}(y)=0.
\end{alignat*}
These two equations together with the boundary conditions require

\begin{equation}
\gamma_{1}\gamma_{2}\chi_{c}=+i\chi_{c},\ \ \gamma_{3}\gamma_{4}\chi_{c}=+i\chi_{c}.\label{eq:S_condition}
\end{equation}
Using the relation $\gamma_{5}=\tau_{3}\sigma_{0}$ and $\gamma_{1}\gamma_{2}\gamma_{3}\gamma_{4}\gamma_{5}=-1$,
we note that $\chi_{c}$ is the eigen state of $\gamma_{5}$ as
\begin{equation}
\gamma_{5}\chi_{c}=+1\chi_{c}.
\end{equation}
Combing with the condition Eq. \ref{eq:S_condition}, we only have
\begin{equation}
\chi_{c}=(1,0,0,0)^{T}.
\end{equation}
Finally, the corner mode is
\begin{equation}
\Psi(x,y)=C\chi_{c}(e^{-\zeta_{1}x}-e^{-\zeta_{2}x})(e^{-\lambda_{1}y}-e^{-\lambda_{2}y}),
\end{equation}
where $C$ is a normalization factor, $\zeta_{1,2}=\frac{1\pm\sqrt{1-4m_{x}b}}{2b}$,
and $\lambda_{1,2}=\frac{1\pm\sqrt{1-4m_{y}b}}{2b}$.

For a square sample, the other two neighboring corner modes are eigenstates
of $\gamma_{5}$ with eigenvalue  $-1$ as $\gamma_{5}\chi_{c}=-1\chi_{c}.$

\section{Self-consistent Born approximation analysis }

In this section, we provide some details for the self-consistent Born
approximation (SCBA) analysis. We start from the clean Hamiltonian,
as Eq. (1) in the main text,

\begin{alignat}{1}
H_{q}({\bf k}) & =t\sin k_{y}\gamma_{1} +[t_{y}+t\cos k_{y}]\gamma_{2}\\
 &+t\sin k_{x}\gamma_{3}+[t_{x}+t\cos k_{x}]\gamma_{4}\nonumber \label{eq:H_q}
\end{alignat}
where the gamma matrices are $\gamma_{j}\equiv-\tau_{2}\sigma_{j}$
{(}$j=1,2,3${)} and $\gamma_{4}\equiv\tau_{1}\sigma_{0}$ with $\tau$
and $\sigma$ both being Pauli matrices but for different degrees of freedom.
The model parameters $t_{x,y}$ and $t$ are defined
the same as in the main text. In our case we consider the disorder with
the form $V({\bf r})\gamma_{4}$, the full Hamiltonian is
\begin{equation}
\mathcal{H}=H_{q}({\bf k})+V({\bf r})\gamma_{4},
\end{equation}
where the potential $V({\bf r})$ distributes uniformly within the
interval $[-W/2,W/2]$.

Then the effects of disorder can be accounted in terms of the self-energy
defined as

\begin{equation}
\frac{1}{E_{F}-H_{q}({\bf k})-\text{\ensuremath{\Sigma(E_{F})}}}=\langle\frac{1}{E_{F}-\mathcal{H}}\rangle,
\end{equation}
where $\langle...\rangle$ denotes the average over all disorder configurations.
Consider the symmetry of the Hamiltonian and Brillouin zone, the self-energy
is simplified to a from

\begin{equation}
\Sigma(E_{F})=\Sigma_{4}\gamma_{4}+\Sigma_{2}\gamma_{2}+\Sigma_{0}I_{4\times4}.
\end{equation}
$E_{F}$ is Fermi energy and it is set as zero in our following discussion.

The key is to get self-energy self-consistently, which will renormalize
the original Hamiltonian. Note that the real part of self-energy renormalizes
parameters of original Hamiltonian, while the its imaginary parts
give band broadening and life time of quasiparticles. In the self-consistent
Born approximation, the self-energy $\Sigma$ is given by the integral
equation

\begin{equation}
\Sigma=\frac{W^{2}}{48\pi^{2}}\int_{BZ}d^{2}{\bf k}\gamma_{4}\frac{1}{E_{F}+i\delta-H_{q}({\bf k})-\Sigma}\gamma_{4},\label{eq:SCBA}
\end{equation}
where the integral runs over the first Brillouin zone, and $\delta$
is an infinitesimal positive number. From Eq.~\eqref{eq:SCBA}, there
are explicitly three coupled self-consistent integral equations
\begin{alignat}{1}
\Sigma_{0} & =\frac{W^{2}}{12}\frac{a^{2}}{4\pi^{2}}\int_{BZ}d^{2}{\bf k}\frac{i\delta-\Sigma_{0}}{(i\delta-\Sigma_{0})^{2}-E^{2}(k)},\nonumber \\
\Sigma_{4} & =\frac{W^{2}}{12}\frac{a^{2}}{4\pi^{2}}\int_{BZ}d^{2}{\bf k}\frac{t_{x}+t\cos k_{x}+\Sigma_{4}}{(i\delta-\Sigma_{0})^{2}-E^{2}(k)},\nonumber \\
\Sigma_{2} & =-\frac{W^{2}}{12}\frac{a^{2}}{4\pi^{2}}\int_{BZ}d^{2}{\bf k}\frac{t_{y}+t\cos k_{y}+\Sigma_{2}}{(i\delta-\Sigma_{0})^{2}-E^{2}(k)},\label{eq:SCBA_3integrals}
\end{alignat}
where $E^{2}(k)=(\Sigma_{4}+t_{x})^{2}+2(\Sigma_{4}+t_{x})t\cos k_{x}+t^{2}+(\Sigma_{2}+t_{y})^{2}+2(\Sigma_{2}+t_{y})t\cos k_{y}+t^{2}.$

After obtaining the self-energy $\Sigma$, the mass terms
are renormalized to
\begin{alignat}{1}
\bar{t}_{x} & =t_{x}+\mathrm{Re}\Sigma_{4},\nonumber \\
\bar{t}_{y} & =t_{y}+\mathrm{Re}\Sigma_{2}.
\end{alignat}
Then the phase boundary of topological quadrupole insulators is reset
at
\begin{equation}
|\bar{t}_{x}|=|\bar{t}_{y}|=1.
\end{equation}

\begin{figure}
\includegraphics[width=0.8\linewidth]{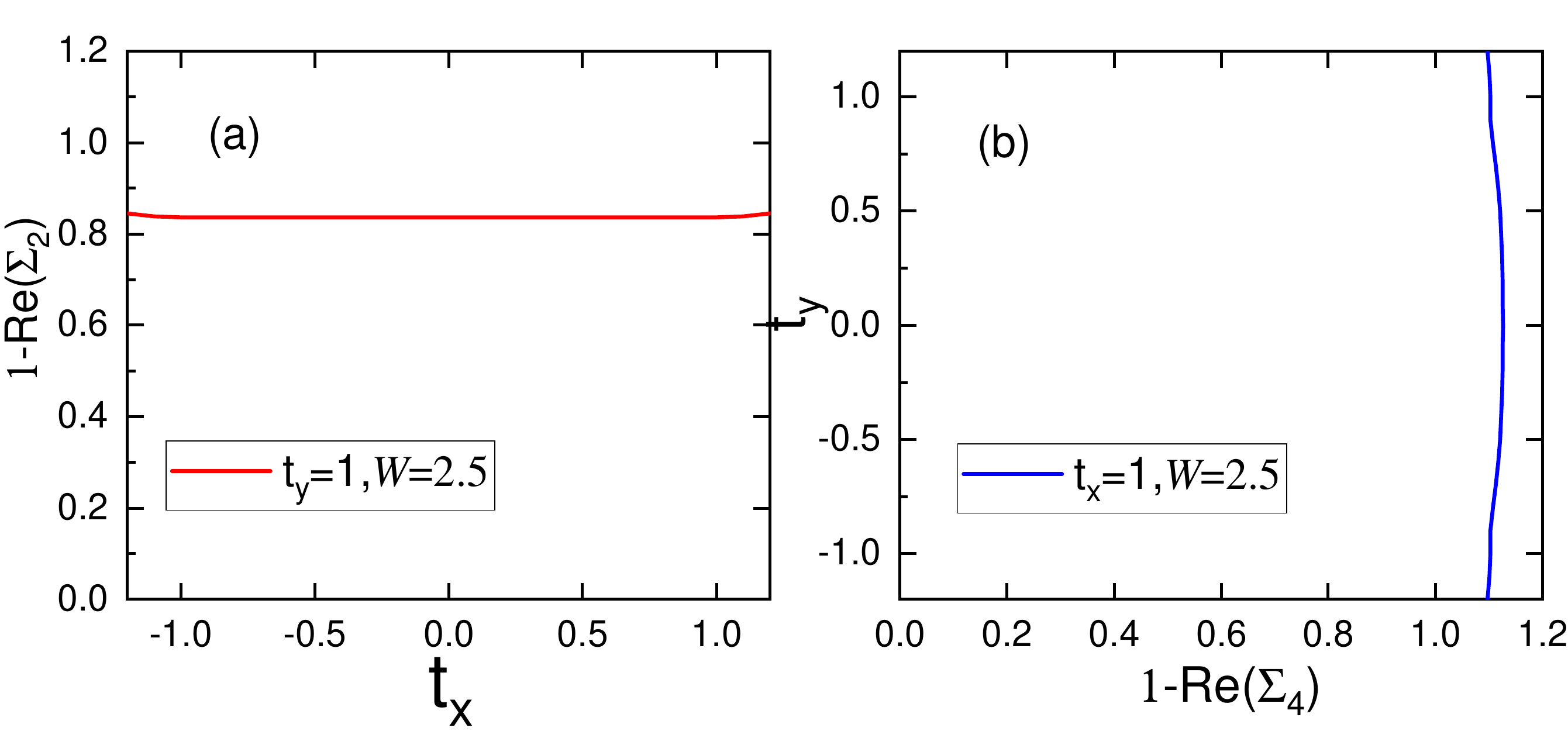}

\caption{Self-consistent Born analysis of phase boundaries. (a) Insensitivity
of modification $\mathrm{Re}[\Sigma_{2}]$ as changing parameter $t_{x}$.
(b) Insensitivity of modification $\mathrm{Re}[\Sigma_{4}]$ as changing
parameter $t_{y}$. \label{fig:S_SCBA}}
\end{figure}

As discussed in the main text on the phase diagram Fig. 1(b), the
upper horizontal phase boundary is nearly flat thus it shows little
dependence on $t_{x}$. Indeed, this observation is true, which is
verified by the self-consistent Born approximation, as shown in Fig.
\ref{fig:S_SCBA}(a). And we can take the similar approach to the
phase boundary along $t_{y}$ direction. From the integral equations,
one can find it also shows little dependence on $t_{y}$.

\section{Signatures of disorder induced nontrivial phase}

In this section, we show explicitly the corner charges and zero-energy
modes in the disorder-induced topological Anderson phase in electric quadrupole insulators. Corresponding
to the phase diagram Fig. 1(b) in the main text, we choose two scenarios to illustrate our points:
$t_{x}=0.9$ and $t_{x}=1.05$.

\begin{figure}
\includegraphics[width=0.8\linewidth]{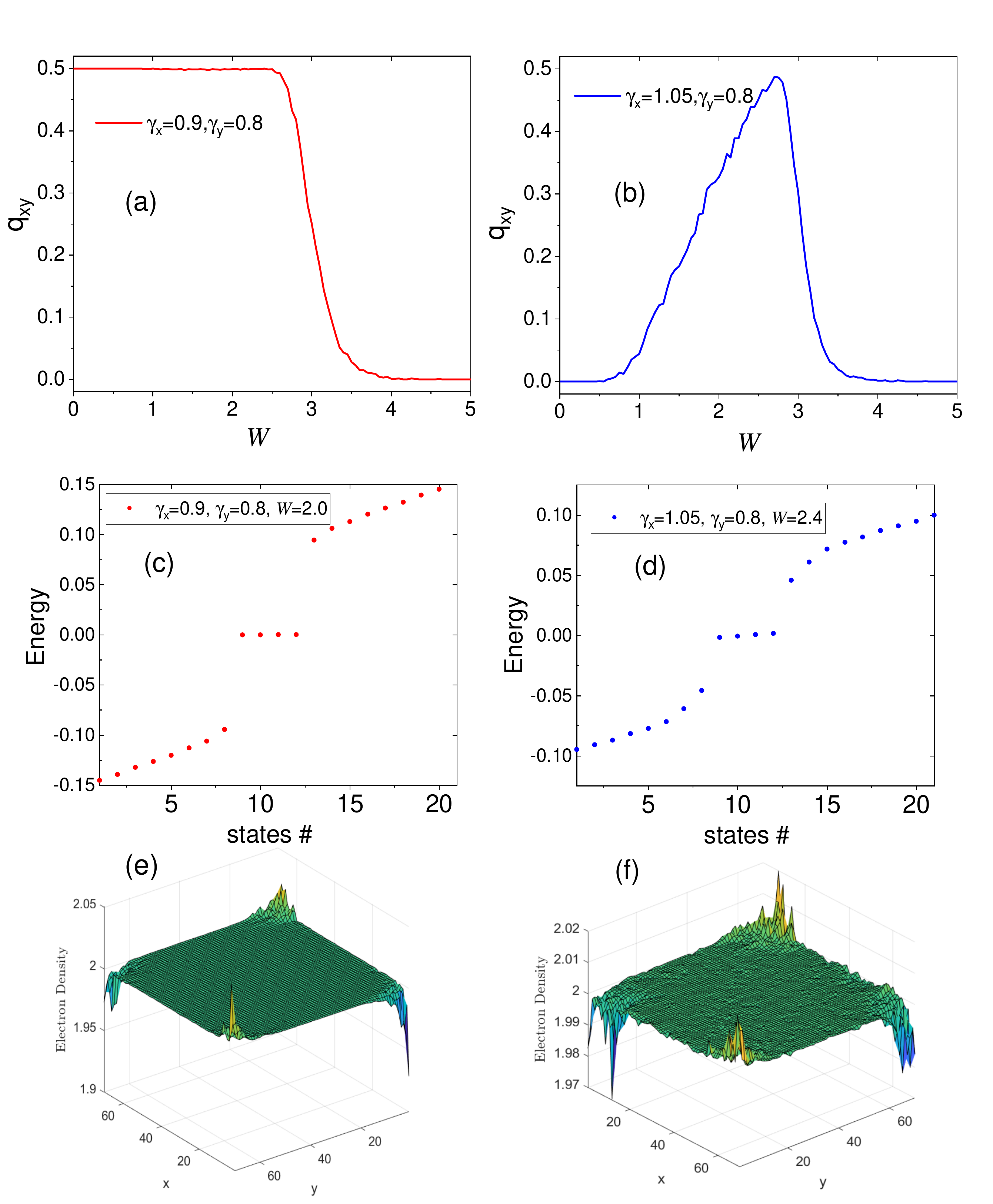}
\caption{Signature of disorder-driven nontrivial phase in electric quadrupole insulators. (a,b) Evolution
of quadrupole moments $q_{xy}$ with disorder strength $W$. Periodic
boundary condition is taken and 1000 disorder configurations are made.
(c,d) The energy modes near the zero energy. (e,f) Electron charge
density corresponds to (c,d), respectively; the system size is $L_{x}\times L_{y}=70\times70$,
and open boundary condition is taken. Here, only one disorder configuration
is taken in (e,f). \label{fig:S_Ingapstates}}
\end{figure}

For the first scenario, $q_{xy}=1/2$ for $W=0$. In the limit $W\rightarrow\infty$,
$W\gamma_{4}$ term dominates and hence $q_{xy}=0$. As such, there
must exist a topological phase transition as increasing the disorder
strength $W$. The varying of $q_{xy}$ as function $W$ is shown
in Fig. \ref{fig:S_Ingapstates}(a), where one finds that $q_{xy}$
stays quantized and has no fluctuations even for quite large $W$(compared
to edge band gap, see the section S7), and no disorder average is necessary.
At this moment, the zero-energy modes (see Fig. \ref{fig:S_Ingapstates}(c))
and quantized corner changes (see Fig. \ref{fig:S_Ingapstates}(e))
are robust against disorders. Around the critical point $W_{c}\simeq3$,
the $q_{xy}$ suddenly switch from $q_{xy}=1/2$ to $q_{xy}=0$, accompanied
by strong fluctuations during this phase transition.

For the second scenario, $q_{xy}=0$ for $W=0$. Similarly, in the
limit $W\rightarrow\infty$, $q_{xy}=0$. The difference is that by
increasing the strength, the disorder can gradually drive $q_{xy}$
away from zero and reach the value near one-half, then decay to zero
finally, see Fig. \ref{fig:S_Ingapstates}(b). This process indicates
a topological phase transition driven by disorder from trivial phase
to nontrivial phase. Finally the system is back to trivial phase again. The appearance
of the nontrivial phase induced by disorders is evidenced by the zero-energy
modes (see Fig. \ref{fig:S_Ingapstates}(d)). The corresponding corner
charge (see Fig. \ref{fig:S_Ingapstates}(f)) approaches to quantized values as the system size is large
enough.

\section{The real part of quadrupole moments formula}

In this section, we use information obtained from the real part of
quadrupole moments formula to detect phase boundaries. From Resta's
construction \cite{Resta98prl,Resta99prl}, besides the fact that imaginary
part of $\mathrm{det}(U^{\dagger}\hat{P}U)$ gives rise the polarization,
its real part also provides useful information, i.e., the localization
length of ground states (note that $\hat{P}=\exp[i2\pi\hat{x}/L_{x}]$).
Taking a close analogy, the real part of $\mathrm{det}(U^{\dagger}\hat{Q}U)$
should also provide invaluable information. Here we define the ``localization
length'' of ground states in electric quadrupole insulators as \cite{Resta99prl}
\begin{equation}
\xi=-\mathrm{Relog\mathrm{det}(U^{\dagger}\hat{Q}U)}.\label{eq:Localization length}
\end{equation}

Let us first verify that $\xi$ calculated from Eq.~\eqref{eq:Localization length}
can be used to detect phase transition. In Fig. \ref{fig:realpart}(a),
we compare the scaling behavior of $\xi$ for different sample size
in the clean system. Note that $\xi$ is scaled by factor of $\mathrm{log}L$
in Fig. \ref{fig:realpart} by considering finite size scaling \cite{Kang19prb}.
Two features in Fig. \ref{fig:realpart}(a) are obvious. First, a
peak stands approximately at $t_{x}=1$, which is the phase boundary
as we know from phase diagram. When $t_{x}<1$, the system is in topological
nontrivial phase, and when $t_{x}>1$, the system is trivial. Second,
the peak becomes higher and approaches closer to $t_{x}=1$ as system
size is enlarged. One can expect that the position of the peak finally
locates at $t_{x}=1$, and its peak values of $\xi$ approaches to
infinity in the thermodynamic limit. Thus, based on this peak position,
we may identify the phase boundary.

\begin{figure}
\includegraphics[width=0.8\linewidth]{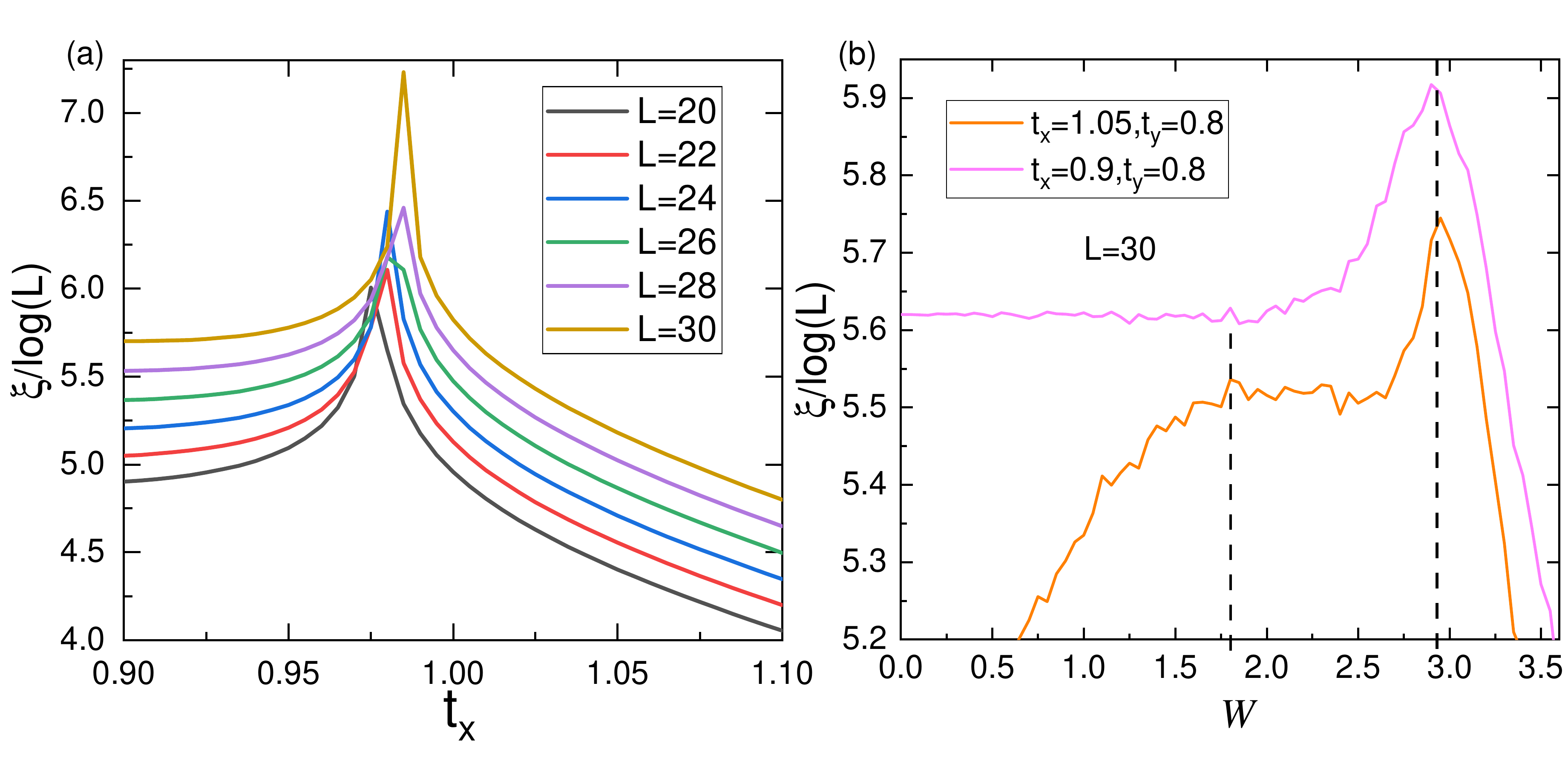}

\caption{Phase transition of electric quadrupole insulators indicated by the
localization length. (a) $\xi$ as function of $t_{x}$ for different
system size in the clean case. Here $t_{y}=0.75$. (b) $\xi$ as
function of disorder strength $W$. The system size is set as $L\equiv L_{x}=L_{y}=30,$
and $1000$ disorder configurations are taken. \label{fig:realpart}}
\end{figure}

Then let us turn to the case of disordered systems, as shown in Fig.
\ref{fig:realpart}(b). Here two typical scenarios of phase transition
are exemplified to illustrate our points. In the first scenario {[}see
pink line in Fig. \ref{fig:realpart}(b){]}, disorder drives the system
directly from topological quadrupole insulator to a trivial insulator.
In this case, $\xi$ nearly keeps a constant with negligible fluctuations,
like a plateau, for small $W$, and one\textcolor{blue}{{} }may treat
this robustness feature of $\xi$ as a signature of topological nontrivial
phase. As the increasing of $W$, $\xi$ strikes a peak at critical
value $W_{c1}=2.9$ at which the topological phase transition occurs.
While in the second scenario {[}see orange line in Fig. \ref{fig:realpart}(b){]},
the system starts from trivial phase and experiences a disorder-induced topological
Anderson phase before entering the trivial phase
under sufficiently strong disorder. As the increasing of $W$ in this
case, $\xi$ first grows to a plateau then also experiences a peak.
Actually, the first phase transition from trivial to nontrivial phase
is vaguely captured by a small peak {[}see the left dashed line{]},
and the narrow plateau afterwards solids the existence of topological
nontrivial phase. While the followed peak of $\xi$ clearly indicates
the phase boundary. Concluding from both scenarios, the phase boundary
between trivial and nontrivial phases locates at $W_{c2}\simeq2.96$,
which is consistent with the phase diagram in Fig. 1 in the main text.
As we can see, the calculation of $\xi$ works as a complement to
phase boundary in Fig. 1 in the main text. If we consider a third scenario,
say $t_{x}=1.5$ in Fig. \ref{fig:S_SCBA}, it is found that $\xi$
grows all the way following $W$ from $0$ to $4$ and shows no sign
of peak, which is also consistent with the phase diagram.

\section{Finite size scaling result}
\begin{figure}
\includegraphics[scale=0.6]{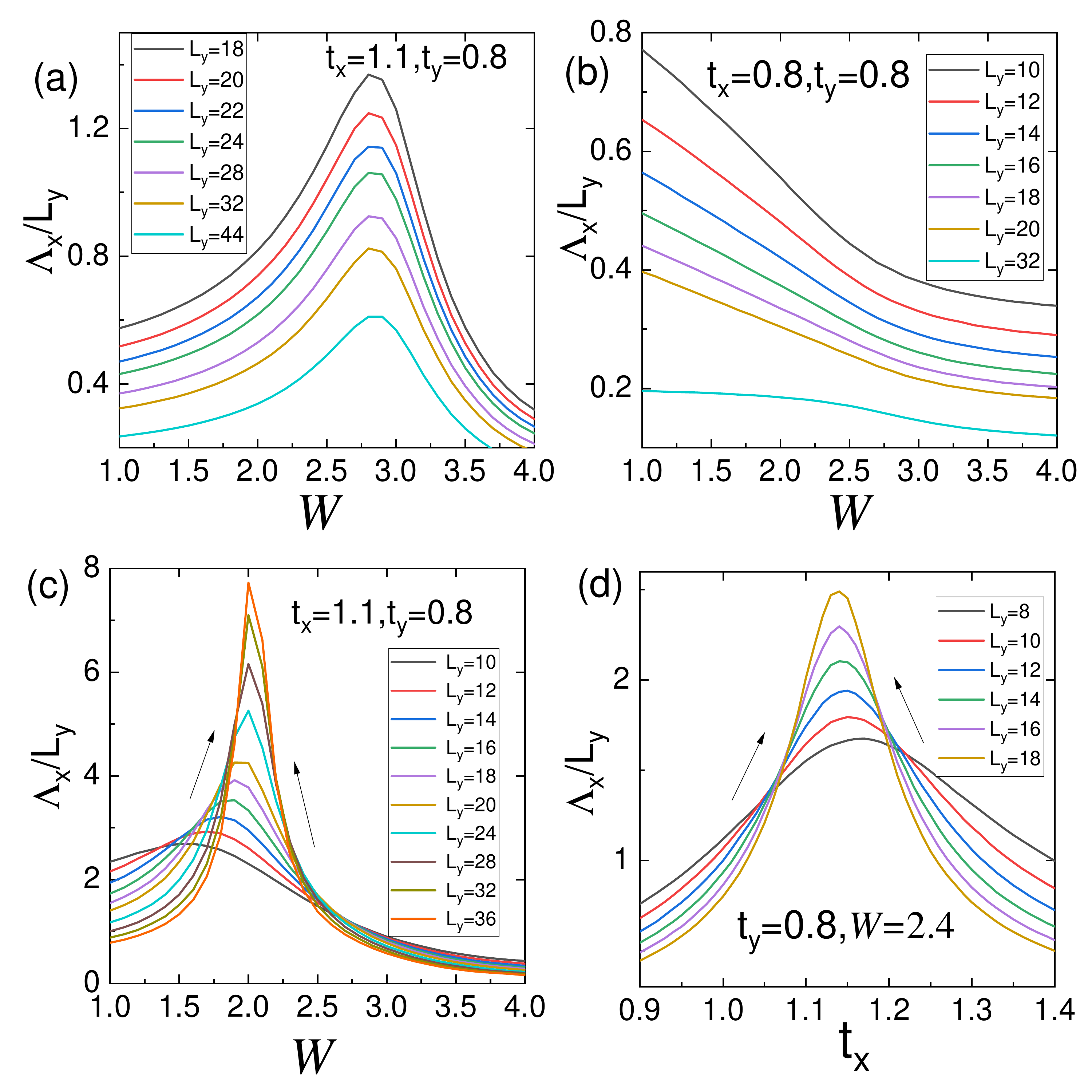}

\caption{Dimensionless localization length for different cases corresponding
to the phase diagram Fig. 1 in the main text. The periodic boundary condition
is only applied to (a). \label{fig:S_scaling}}
\end{figure}
In this section, we explore the signatures of delocalized states at the
sample boundaries as consequence of disorder-driven topological phase
transitions.
Consider a long ribbon ($2\times10^6$ in units of lattice constant) along the $x$ direction. The localization length is calculated using
the transfer matrix method. For the single-parameter scaling, we plot
the dimensionless localization length $\Lambda/L_{y}$ as functions of disorder strength $W$ (or parameter $t_x$) in
Fig. \ref{fig:S_scaling}. As we discussed in the main text, since the bulk is always trivial, finite size scaling with periodic boundary condition (along the transverse direction of the ribbon) cannot reveal the topological phase transitions of the electric quadrupole insulators,
as shown in Fig. \ref{fig:S_scaling}(a). Then we try the case with
open boundary condition.
From the finite size scaling theory,
the dimensionless localization length $\Lambda_{x}/L_{y}$ becomes
scale free near the critical point of metal-insulator transition.
For the case of $t_{x}=t_{y}=0.8$, a phase transition occurs
during the increasing of $W$(see Fig. 1 in the main text), while
the finite size scaling result in Fig. \ref{fig:S_scaling}(b) shows
no critical point, and thus the bulk and boundaries along the $x$
direction show insulating behavior. If we tune the parameters to $t_{x}=1.1,t_{y}=0.8$,
it seems that two ``critical points'' $W_{c1}$ and $W_{c2}$ occur during
the increasing of $W$, and the region between $W_{c1}$ and $W_{c2}$
exhibit a metallic behavior. While this signature is ``false'' since
as increasing the width $L_{y}$ further, the two ``critical points'' approach
to phase boundary point $W_{c}$ simultaneously, at which the localization
length $\Lambda_{x}$ diverges (see Fig. \ref{fig:S_scaling}(c)).
To verify this point, we take $t_{x}$ instead of $W$ as the scaling
parameter in Fig. \ref{fig:S_scaling}(d). Similarly, the two ``critical points''  will approach to the phase boundary point as increasing the
width $L_{y}$.

\section{The edge Hamiltonian based on lattice model}

In this section, we derive the exact edge Hamiltonian based on the lattice model.  Let us
first observe the energy band of an infinitely long ribbon along the $x$ direction
but with finite width in the $y$ direction, as shown in Fig. \ref{fig:S_edgegap}(a).
The red lines indicates the edge bands whose wave function localized
at the edges. It is interesting to find that these edge bands, unlike
other bulk bands, do not shift as varying $t_{y}$ as long as $|t_{y}|<1$
{[}see Fig. \ref{fig:S_edgegap}(b){]}. Under the condition $|t_{y}|\text{<1},$
the edge band gap is
\begin{equation}
E_{gx}=2|1-t_{x}|.
\end{equation}

In the following we derive the exact form of the edge Hamiltonian, which will give the edge band gap naturally.
Assume the ribbon width is $N_{y}$ and take open boundary condition at $y$ direction.
By partial Fourier transformation of the lattice Hamiltonian along
the $x$ direction, it reads
\begin{equation}
H=\sum_{k_{x}}\left[\sum_{R_{j}=1}^{N_{y}}\Psi_{k_{x},R_{j}}^{\dagger}h_{j,j}(k_{x})\Psi_{k_{x},R_{j}}+\sum_{R_{j}=1}^{N_{y}-1}\Psi_{k_{x},R_{j}}^{\dagger}h_{j,j+1}(k_{x})\Psi_{k_{x},R_{j}+1}+h.c.\right],\label{eq:DSSH}
\end{equation}
where
\begin{equation}
h_{j,j}(k_{x})=\left(\begin{array}{cccc}
0 & 0 & t_{x}+te^{ik_{x}} & t_{y}\\
0 & 0 & -t_{y} & t_{x}+te^{-ik_{x}}\\
t_{x}+te^{-ik_{x}} & -t_{y} & 0 & 0\\
t_{y} & t_{x}+te^{ik_{x}} & 0 & 0
\end{array}\right),h_{j,j+1}(k_{x})=\left(\begin{array}{cccc}
0 & 0 & 0 & t\\
0 & 0 & 0 & 0\\
0 & -t & 0 & 0\\
0 & 0 & 0 & 0
\end{array}\right),
\end{equation}
and $\Psi_{k_{x},R_{j}}=[C_{k_{x},R_{j},1},C_{k_{x},R_{j},2},C_{k_{x},R_{j},3},C_{k_{x},R_{j},4}]^{T}.$

We note that Eq.~\eqref{eq:DSSH} describes a four-band Su-Schrieffer-Heeger
(SSH) chain along the $y$ direction with open boundary. In the case
of $t=1$, it is topologically nontrivial when $|t_{y}|<1$. Thus
it gives the end states as $E_{\mathrm{edge}}(k_{x},t_{x})$, which
is exactly the edge band indicated by red line in Fig. \ref{fig:S_edgegap}(a).
Usually, it is hard to get the $E_{\mathrm{edge}}(k_{x},t_{x})$ for
general $t_{y}$, but let us focus on the simplest case $t_{y}=0,$
at which the four-band SSH chain decouples to a block diagonal form

\begin{equation}
H=\mathrm{diag}(H_{\mathrm{SSH,}x},B,B,\cdots,B,H_{\mathrm{SSH,}x}),
\end{equation}
where
\begin{equation}
H_{\mathrm{SSH,}x}=\left(\begin{array}{cc}
0 & t_{x}+te^{-ik_{x}}\\
t_{x}+te^{ik_{x}} & 0
\end{array}\right),B=\left(\begin{array}{cccc}
0 & t_{x}+te^{ik_{x}} & 0 & t\\
t_{x}+te^{-ik_{x}} & 0 & -t & 0\\
0 & -t & 0 & t_{x}+te^{-ik_{x}}\\
t & 0 & t_{x}+te^{ik_{x}} & 0
\end{array}\right).
\end{equation}
At this point, we totally dimerized the Hamiltonian, and the end modes
are encoded in the two matrices $H_{\mathrm{SSH},x}$ at the ends
of the chain, and we find that $H_{\mathrm{SSH},x}$ is exactly the
SSH model. By diagonalizing
$H_{\mathrm{SSH},x}$, such that
\begin{equation}
E_{\mathrm{edge}}(k_{x},t_{x})=\pm\sqrt{t_{x}^{2}+t^{2}+2t_{x}t\cos k_{x}},
\end{equation}
which gives the edge bands as denoted by the red lines in Fig. \ref{fig:S_edgegap}(a).

\begin{figure}
\includegraphics[width=0.8\linewidth]{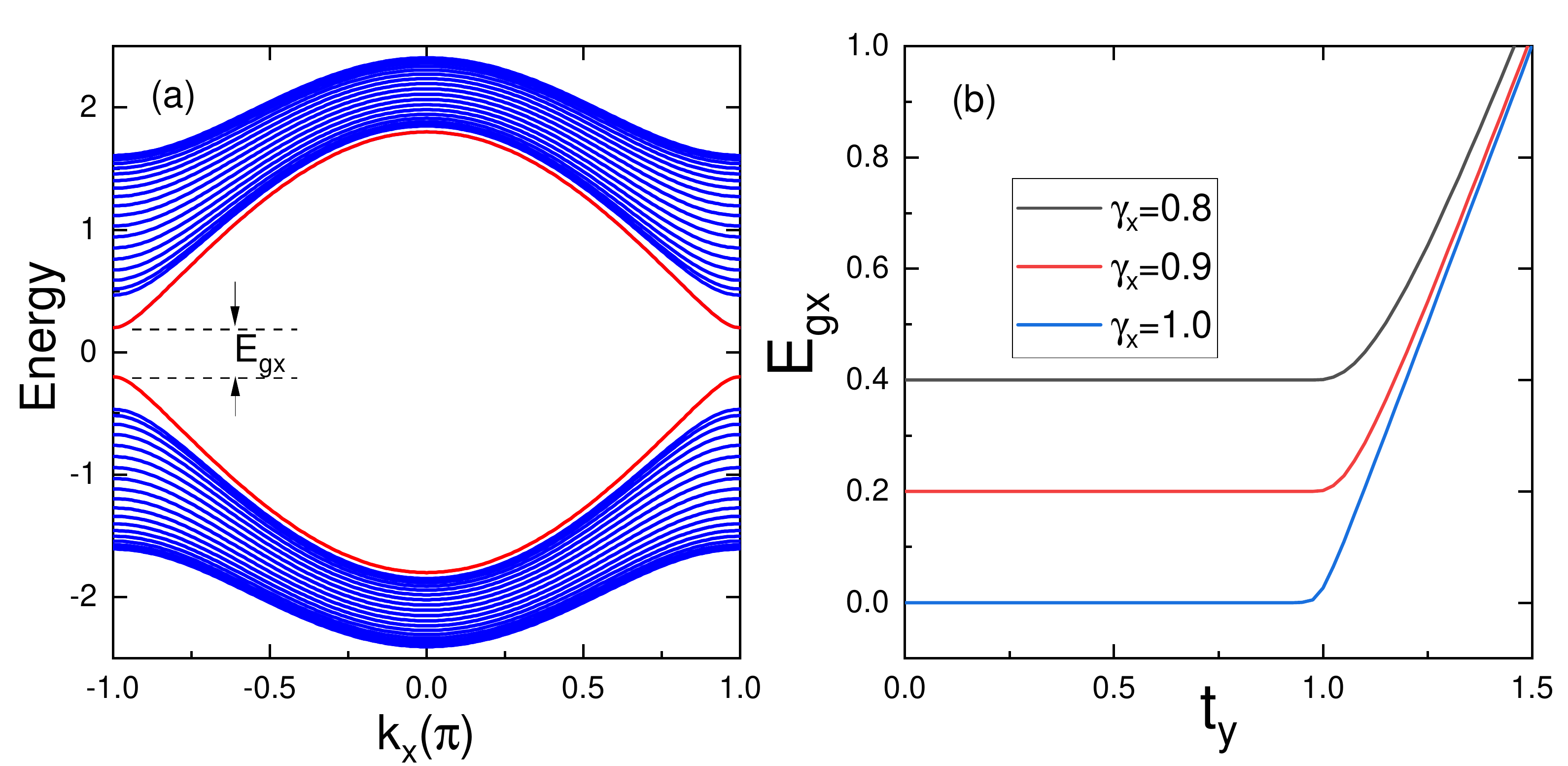}

\caption{The edge band gap. (a) Energy spectrum for $t_{x}=0.8,t_{y}=0.6$.
(b) Edge band gap as function of $t_{y}$. \label{fig:S_edgegap}}
\end{figure}

For $t_{y}\neq0$, the Hamiltonian Eq.~\eqref{eq:DSSH} cannot be
totally dimerized thus is difficult to solve. Now we consider a semi-infinite
plane with $y\geq0$ and keep $k_{x}$ good quantum number. Assume
the edge states wave function
\begin{equation}
\psi(R_{i}^{x},R_{j}^{y})=\frac{1}{\sqrt{N_{y}}}e^{ik_{x}R_{i}^{x}}\phi(R_{j}^{y}),
\end{equation}
where $\phi(R_{j}^{y})$ is the spatial part along the $y$ direction
that contains the decay factors. By projecting out the degrees of freedom
$R_{j}^{y}$, we have an effective lattice model on the $x$ direction.
The hopping integrals between sites $R_{i,A}^{x}$ and $R_{i,B}^{x}$
is obtained as

\begin{equation}
\sum_{R_{j}^{y}}\psi^{*}(R_{i,A}^{x},R_{j}^{y})t_{x}\psi(R_{i,B}^{x},R_{j}^{y})=t_{x},
\end{equation}
and the hopping integrals between sites $R_{i,B}^{x}$ and $R_{i+1,A}^{x}$
is obtained as
\begin{equation}
\sum_{R_{j}^{y}}\psi^{*}(R_{i,B}^{x},R_{j}^{y})t\psi(R_{i+1,A}^{x},R_{j}^{y})=te^{ik_{x}}.
\end{equation}
Thus the effective lattice model on the $x$ direction is still the same
SSH model as
\begin{equation}
H_{\mathrm{SSH},x}=\left(\begin{array}{cc}
0 & t_{x}+te^{-ik_{x}}\\
t_{x}+te^{ik_{x}} & 0
\end{array}\right)
\end{equation}
under the basis $(C_{k_{x},A},C_{k_{x},B})^{T}$ where $C_{k_{x},A/B}$
is the annihilation operator. It is numerically verified true that
the edge bands should remain the same no matter what $t_{y}$ is when
$|t_{y}|<1$. At $k_{x}=\pi,$ it gives the edge band gap $E_{gx}$
as stated before. For $|t_{y}|>1$, no edge bands exist.

If we focus on the edge band along the $y$ direction, we have the
similar result as
\begin{equation}
H_{\mathrm{SSH},y}=\left(\begin{array}{cc}
0 & t_{y}+te^{-ik_{y}}\\
t_{y}+te^{ik_{y}} & 0
\end{array}\right),
\end{equation}
and corresponding edge bands
\begin{equation}
E_{\mathrm{edge}}(k_{y},t_{y})=\pm\sqrt{t_{y}^{2}+t^{2}+2t_{y}t\cos k_{y}}.
\end{equation}

\end{widetext}

\end{document}